\documentstyle[graphicx,amsmath]{elsart}

\begin{document}
\title{"Qualms" from Lavenda, cond-mat/0311270}
\author{D.H.E. Gross}\vspace{-0.5cm}*
\address{Hahn-Meitner-Institut
  Berlin, Bereich Theoretische Physik,Glienickerstr.100\\ 14109
  Berlin, and Freie Universit{\"a}t Berlin, Fachbereich
  Physik, Germany, and
Universit\`a di Firenze and INFN, Sezione di Firenze, via Sansone
1, 50019 Sesto F.no (Firenze), Italy. \today} \maketitle Abstract:
The smoke (German: Qualm) produced by Lavenda obscures the simple
and fundamental physics of non-extensive systems as proposed in
\cite{gross185}.
\\~\\
The paper \cite{gross185} addresses phase-transitions in systems with a
fixed, finite, volume, here the Potts-($q=3$)-lattice gas on a $50\times
50$ lattice, details in the cited papers \cite{gross174,gross173}.
Therefore, the volume is irrelevant for the transition. We have a system
with the {\em two} control-parameters $E,N$. For this case Lavenda makes
several wrong statements:
\begin{enumerate}
\item {\bf "The author discards the fundamental properties of concavity".} Yes, this is the
essence of my new extended formulation of thermo-statistics: At
phase-separation the entropy [$S(E)$ not $S(E)/N$(!)] of any
system is convex (bimodal). Therefore, in conventional
thermo-statistics one gets in the thermodynamic limit Yang-Lee
singularities. These are certainly not helpful to give a detailed
insight into this most interesting physics: The original objects
of Thermodynamics, steam engines, work exactly at
phase-separation. Sailing on the surface of the sea would,
presumably be impossible if there were no phase-separation. In hot
nuclei this is the region where all the interesting fragmentation
phenomena occur, from nucleon evaporation, to fission, to
multifragmentation, and finally to vaporization.
\item {\bf "The curvature eigenvalues $\lambda_1\ge\lambda_2$ are devoid of any physical meaning."}
On the contrary: for a totally concave entropy surface $S(E,N)$ the
condition $\lambda_1<0$ guarantees that Lavenda's equation (4) for any
$u,v$ at fixed volume $V$ is fulfilled:
\begin{equation}
S_{EE}\times u^2 +2S_{EN}\times uv+S_{NN}\times v^2\le 0 \label{eq1}
\end{equation}
This can also be seen by starting from Lavenda's equation for the
eigen-curvatures :
\begin{equation}
\lambda_{1,2}=\frac{1}{2}(S_{NN}+S_{EE})\pm \frac{1}{2}\sqrt{(S_{NN}-S_{EE})^2+4S^2_{EN}}
\end{equation}
If both eigenvalues are $<0$ then evidently both $S_{NN}$ and $S_{EE}$ must be negative.
Therefore, the single condition $\lambda_1<0$ suffices to guarantee relation (\ref{eq1}),
and of course also Lavenda's conditions (2+3).
\item {\bf "The inverse temperature $\beta$ has no meaning in the microcanonical ensemble".}
Yes, of course, however, what was discussed in my paper
\cite{gross185} is the weight of a given energy in the Laplace
transform  $E\to T$:
\begin{equation}
Z(T,\mu,V)=\int_0^{\infty}{dE\;e^{S(E)-\beta E}},
\end{equation} with the external parameter $\beta=1/T$.
The weight $e^{S(E)-\beta E}$ has a single maximum in mono-phases,
and  at phase-separation it is bimodal, where it has at least two
sharp maxima separated by the latent heat with a minimum in
between (convexity).
\item {\bf "Extensivity of the entropy is already incorporated in Boltz\-mann--Planck's
principle {\boldmath$S=k\ln W$}"}. First, extensivity and
additivity are two different things. For two independent and
non-interacting systems the microcanonical Boltzmann-Planck
entropy is of course additive, $S=S_1+S_2$. Extensivity means
something else: $S(N)$ scales as $S(N)\propto N$. This is usually
demanded in the thermodynamic limit but is, of course, invalid in
non-extensive situations.
\end{enumerate}

\end{document}